\documentclass{article}
\usepackage[utf8]{inputenc}
\usepackage{array}
\usepackage{graphicx}
\usepackage{amsmath}
\usepackage{amsfonts}
\usepackage{indentfirst}
\title{Macro carbon price prediction with support vector regression and Paris accord targets. }
\author{Jinhui Li}
\date{September 2022}

\begin{document}

\maketitle
\begin{abstract}
     Carbon neutralization is an urgent task in society because of the global warming threat. And carbon trading is an essential market mechanics to solve carbon reduction targets. Macro carbon price prediction is vital in the useful management and decision-making of the carbon market. We focus on the EU carbon market and we choose oil price, coal price, gas price, and DAX index to be the four market factors in predicting carbon price, and also we select carbon emission targets from Paris Accord as the political factor in the carbon market in terms of the macro view of the carbon price prediction. Thus we use these five factors as inputs to predict the future carbon yearly price in 2030 with the support vector regression models. We use grid search and cross validation to guarantee the prediction performance of our models. We believe this model will have great applications in the macro carbon price prediction.
\end{abstract}
\section
{Introduction}

With the rapid growth of the world economy, the problem of environmental pollution has gradually entered people's vision. Among them, global warming is an important aspect of environmental pollution.In the last century,the global average temperature has risen by approximately one degree Celsius,and the rate of increase is accelerating.\cite{1} There is a growing scientific consensus that rising concentrations of carbon dioxide (CO2) and other greenhouse gases, which result from the burning of fossil fuels, are gradually warming the Earth’s climate. Carbon dioxide emissions are the primary cause of global warming.\cite{2}  Carbon emissions trading is a critical strategy to minimize carbon dioxide emissions. Previous studies represent one of the first comprehensive studies of the impact of oil price risk on emerging stock markets. Researchers find strong evidence that oil price risk impacts stock price returns in emerging markets. Results for other risk factors like market risk, total risk, skewness, and kurtosis are also presented.\cite{3} 
 
\subsection{Glasgow Climate Pact}

Since the entry into force of the United Nations Framework Convention on Climate Change (UNFCCC) in 1994 and the Kyoto Protocol in 2005 to the Paris Agreement in 2015, the UN-led global climate governance has undergone more than 24 years of tortuous development. International rules on emissions reduction have evolved from a top-down model under the Kyoto regime to a "bottom-up" model under the Paris regime.\cite{4}

2021, 13 November, the 26th session of the Conference of the Parties to the United Nations Framework Convention on Climate Change (COP26, also known as the "Glasgow Summit") concluded successfully with a consensus on the practical implementation of the Paris Agreement and other related elements, and nearly 200 countries signed the Glasgow Climate Convention after consultations and discussions among delegates. COP26 reached an agreement on the market and non-market rules for carbon trading under Article 6 of the Paris Agreement, potentially freeing up trillions of dollars for forest protection, renewable energy facilities and other projects to combat climate change. At the heart of the agreement are "measures to ensure that carbon credits are not double-counted under each country's emissions targets". Moreover, to ensure that old credits do not flood the market, carbon credits issued before 2013 will not be allowed to be carried forward.

By 14 November, the end of COP26, 151 countries worldwide had joined the net zero pledge, representing 90\% of global GDP and 89\%  of global emissions. At COP26, US President Joe Biden and the President of the European Commission officially launched the Global Methane Emissions Reduction Commitment (GEMEC), also known as the '3030 Commitment', as the signatories agreed to reduce methane emissions by 30\% by 2030 compared to the current levels. During COP26, more than 40 countries, including China, the US, the EU, India, the UK and Australia, launched the Glasgow World Leaders Breakthrough Agenda. This agreement will enable countries and businesses to work together to accelerate the development and deployment of the clean energy technologies and sustainable solutions needed to meet the goals of the Paris Agreement over the next ten years. COP26 climate summit announces an international agreement to accelerate the end of coal use, shifting the goal from 'no new coal' to 'complete phase-out of coal. 

\subsection{Policy options for reducing emissions}

Here are some policy options for reducing emissions. Incentive-based approaches can reduce emissions at a lower cost than more restrictive command-and-control approaches because they provide more flexibility about where and how emission reductions are achieved. The most efficient approaches to reducing emissions of CO2 involve giving businesses and households an economic incentive for such reductions. Such an incentive could be provided in various ways, including a tax on emissions, a cap on the total annual level of emissions combined with a system of tradable emission allowances, or a modified cap-and-trade program that includes features to constrain the cost of emission reductions that would be undertaken to meet the cap.\cite{5}
Because of this, this article separately presents Brent oil, DAX index, coal price, gas price and carbon emission.

\subsection{The advantage of SVR for our forecasting}
This paper uses a support vector regression approach to predict Brent oil, DAX index, coal price, gas price, and carbon emission. SVR is a variation of support vector machines, able to provide regression models for non-linear datasets.
SVR allows us to define how much error is acceptable in our model and will find an appropriate line (or hyperplane in higher dimensions) to fit the data.

\section{Methodology}
\subsection{Support vector regression}

In machine learning, support-vector machines (SVM), also support-vector networks) are supervised learning models with associated learning algorithms that analyze data for classification and regression analysis. When data are unlabelled, supervised learning is not possible, and an unsupervised learning approach is required, which attempts to find natural clustering of the data to groups, and then map new data to these formed groups. The support-vector clustering algorithm applies the statistics of support vectors, developed in the support vector machines algorithm, to categorize unlabeled data.

SVR gives us the flexibility to define how much error is acceptable in our model and will find an appropriate line (or hyperplane in higher dimensions) to fit the data.
SVR has three kernel :linear, polynomial and rbf (radical basis function) 
Which fits our data perfectly as some of our data has linear patterns and some of our data has polynomial/rbf pattern.
SVR can be used with grid search and cross validation to find the best regression model such that it won't underfit.
SVM uses the following binary classifier to classify data$\lbrace(x_i,y_i)\rbrace_{i=1}^n$:
$$f_{\omega,\gamma}(x)=\omega^Tx+\gamma$$
where $\omega$ is called normal vector and $\gamma$ is called interpret, $y_i\in\lbrace-1,1\rbrace$.

A perfect classifier should satisfy for all training samples:
$$f_{\omega,\gamma}(x_i)y_i=(\omega^Tx_i+\gamma)y_i>0,  {\forall}i=1,...,n$$
However, since open set is not easily to handle mathematically, we use
$$(\omega^Tx_i+\gamma)y_i\geq1,  {\forall}i=1,...,n$$
In practice, we choose $\omega$ and $\gamma$  which maximize the separation margin:
$$\min_{i=1,...,n}\lbrace\frac{(\omega^Tx_i+\gamma)y_i}{||\omega||}\rbrace=\frac{1}{||\omega||}$$
Geometrically, the margin is the half distance between two decision boundaries $\omega^Tx+\gamma=1$and $\omega^Tx+\gamma=-1$.
The hard margin SVM:
$\min_{\omega,\gamma}{\frac{1}{2}||\omega||^2}$ subject to
$(\omega^Tx_i+\gamma)y_i\geq1, {\forall}i=1,...,n$
The soft margin SVM relaxes the condition of linear separability with
$$\min_{\omega,\gamma,\xi}{[\frac{1}{2}||\omega||^2+C\sum_{i=1}^n{\xi_i}] }$$
subject to 
$$(\omega^Tx_i+\gamma)y_i\geq1-\xi_i,\xi_i\geq0, {\forall}i=1,...,n.$$
Here $\xi_i$  are called slack variables, $C>0$ is a control parameter. 
SVM is a quadratic optimization problem with the general form
$$\min_{\theta}{[\frac{1}{2}\theta^{T}F\theta+f^T\theta]}$$
subject to ${G\theta}\leq{g}$,

where ${G\theta}\leq{g}$  is an elementwise inequality, F  is positive definite. If F is ill-conditioned (has a very small eigenvalue), we need to add a small positive constant to the diagonal elements of F. 

\subsection{Grid search and cross validation}

The traditional way of performing hyperparameter optimization has been grid search, or a parameter sweep, which is simply an exhaustive search through a manually specified subset of the hyperparameter space of a learning algorithm. A grid search algorithm must be guided by some performance metric, typically measured by cross-validation on the training set or evaluation on a hold-out validation set.

Since the parameter space of a machine learner may include real-valued or unbounded value spaces for certain parameters, manually set bounds and discretization may be necessary before applying grid search. Grid search suffers from the curse of dimensionality, but is often embarrassingly parallel because the hyperparameter  settings it evaluates are typically independent of each other.

Cross-validation calculates the accuracy of the model by separating the data into two different populations, a training set and a testing set. In n-fold cross-validation  the data set is randomly partitioned into n mutually exclusive folds, $T_1$,$T_2$,…,$T_n$, each of approximately equal size. Training and testing are performed n times. Each training set consists of $\frac{n-1}{n}$ of the data set and the remaining $\frac{1}/{n}$ is used as test data. In 10-fold cross-validation, a given data set is partitioned into ten subsets. Out of these ten subsets, nine are used to perform a training fold and a single subset is retained as testing data. This cross-validation process is then repeated ten times (the number of folds). The ten sets of results are then aggregated by averaging to produce a single model estimation. The advantage of ten-fold cross-validation over random subsampling is that all objects are used for both training and testing, and each object is used for testing only once per fold.

\subsection{Practical application}
Firstly, we use grid search with the scoring of cross-validation to search for the best parameters for the 5 input variables. The graph is shown below:

\begin{center}
\begin{tabular}{ | m{10em} | m{10cm}| m{10cm} | } 
  \hline
  Brent oil & SVR(C=42,epsilon = 0.5,gamma= 'scale', kernel = 'rbf') \\ 
  \hline
  DAX index & SVR(C=10, kernel='linear',epsilon= 0.00001,gamma= 'auto') \\ 
  \hline
  coal price & SVR(C=3,kernel='rbf',epsilon= 4,gamma = 'scale') \\ 
  \hline
  gas price   & SVR(C=48,gamma = 'auto',epsilon = 0.5 ,kernel='linear') \\
  \hline 
  carbon emission & SVR(C=3878,epsilon = 0.00001,gamma = 'auto',kernel = 'linear')\\
  \hline 
  carbon emission 2 & SVR(C=3878,epsilon = 0.00001,gamma = 'auto',kernel = 'linear')\\

\end{tabular}
\end{center}

 From the above table, we can see that the grid search chooses either 'rbf' or 'linear' kernel for the data. For example, the Brent oil data has a clearly non-linear pattern, thus using 'rbf' kernel generalizes much better prediction results. See figure 1. The variable 'C' has the greatest impact on the SVR model, and we set the grid search for C from 0 to 4000. The reason for that is for the carbon emission and carbon emission2 model, the score of cross validation keeps getting lower as we increase the value of C, and the cross validation value reaches the lowest point when C=3878.

    After we get all the prediction data from 2022 to 2030 of the input factors, then we attach them to their real data from 2007 to 2021, thus we obtain the full data of the five factors from 2007 to 2030.Now we consider these data as input. then we can use grid search again to pick the best parameters and to predict the carbon price. The multivariable SVR grid search takes lots of calculating power, thus we only fine-tune the parameter C and the kernel. See the table below

\begin{center}
\begin{tabular}{ | m{10em} | m{10cm}| m{10cm} | } 
  \hline
Carbon Price & SVR(C = 0.00010,kernel = 'linear') \\ 
\end{tabular}
\end{center}

The results show that the carbon price will reach 114.7 euro in 2030 if the target carbon emission is 2137554 kt in 2030, and the carbon price will reach 203.0 euro if the carbon emission is 1603165.5kt in 2030. 

\section{Results}

\subsection{Forecast of five inputs through SVR}
   
   The price of carbon trading is influenced by several factors, and there are five factors are discussed as the main influencing factors: (1) oil price; (2) DAX index; (3) coal price; (4) gas price; (5) carbon emission. So before predicting the trend of carbon trading price, predicting the trend of these five main factors is necessary.
  
   SVR gives us the flexibility to define how much error is acceptable in our model and will find an appropriate line (or hyperplane in higher dimensions) to fit the data. It can be used with grid search and cross validation to find the best line.
  
   In order to accurately predict the trend of these five factors, we downloaded data from previous decades, and by expanding the amount of data and applying the SVR model, we were able to make the predicted results more consistent with the real level and more accurate.

\subsection{Oil price}

   From the moment oil is discovered, its price is influenced by many factors. For example, political policies, wars, the economy and the discovery of new oil sources can all cause fluctuations in the price of oil. In particular, around 2008, oil prices continued to climb due to the war in Iraq, OPEC production cuts, and the rapid growth in Chinese demand for oil.
 
   It can also be seen in Figure 2 that although oil prices have been declining year after year since 1980, they began to rise again near 2000 in a fluctuating manner. Oil price currently experiences a decline and according to SVR's forecast results, this downward trend will last, and oil price will go through a trend of decreasing and then stabilizing gradually, and by 2030, oil prices will be at 35 euro per barrel. 
   \begin{figure}[h!]
    \centering
    \includegraphics{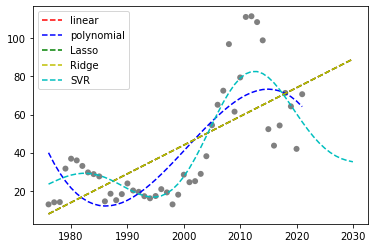}
    \caption{Forecast of oil prices through SVR}
\end{figure}
  
  \subsection{Dax index}
  
   The DAX—also known as the Deutscher Aktien Index or the GER40 - is a stock index representing the 40 largest and most liquid German companies traded on the Frankfurt Stock Exchange (Chen, J). The DAX index, which tracks 40 large, actively traded German companies, is considered by many analysts to be a measure of the health of the German economy. And the companies listed in the DAX index are multinational companies that have a significant impact on the global economy. 

   As can be seen in Figure 3, since 1990, the general trend of the DAX index has been upward, albeit with minor fluctuations. Despite the changing external influences, the rate of increase of the DAX index has always remained in a stable range, and the graph also shows that its growth is basically linear. The DAX index was around 2,000 in 1990, while in 2021 the DAX index has grown to nearly 12,500. So according to the SVR model, the DAX index in 2030 is predicted to reach nearly 15,000.
\begin{figure}[h!]
    \centering
    \includegraphics{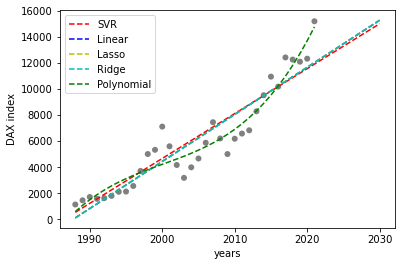}
    \caption{Forecast of DAX index through SVR}
\end{figure}

 \subsection{Coal price}
 
   The price of carbon is influenced by policy factors and the development of other energy sources. In Figure 4 it can be seen that the price of carbon is already the decreasing gradually, compared to 100 euro per ton in 2010, the price of carbon in 2020 has been reduced by half.

   However, in recent years, the international economy and the Russian-Ukrainian conflict have caused some fluctuations in carbon prices beyond expectations. After excluding the disturbance of anomalous data, using the SVR model, we can predict that the carbon price will remain stable in general during 2022-2030, and the carbon price will be around 86 euro per ton in 2030.
\begin{figure}[h!]
    \centering
    \includegraphics{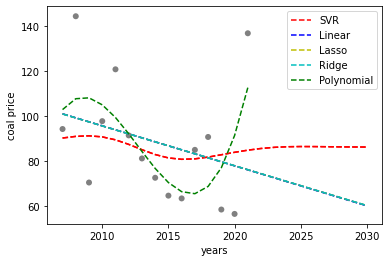}
    \caption{Forecast of coal price through SVR}
\end{figure}

\subsection{Gas price}

   Similar to the DAX index, gas price increases gradually and linearly with the year. Although Russian restrictions on gas exports to European countries following the outbreak of the Russia-Ukraine conflict have led to a surge in gas prices, overall, the price of gas would grow to four times its 2000 level by 2030 in an unrestricted scenario, around 84 euro per unit.
\begin{figure}[h!]
    \centering
    \includegraphics{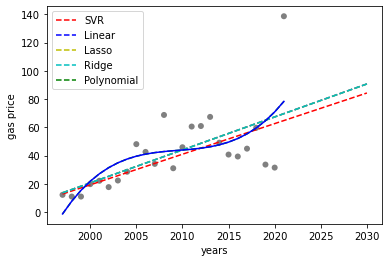}
    \caption{Forecast of the gas price through SVR}
\end{figure}

\subsection{Carbon emission}

   The price of carbon trading is directly linked to carbon emissions, so carbon emissions will be considered as an important influence factor in the discussion. Since the greenhouse effect caused by carbon emissions is contributing to global warming, companies are reducing carbon emissions or capturing carbon dioxide at source as much as possible under the pressure of national policies and models such as carbon trading.

   Figure 6 shows the prediction of carbon emissions using the SVR model under two different targets, and it can be seen that although the targets are different, the carbon emissions show a trend of decreasing year by year. Among them, if the carbon emission is 2137554 kt in 2030, the carbon emission will be reduced to about 2200000 kt in 2030, while if the carbon emission is 1603165.5kt in 2030, the carbon emission will be reduced to about 2250000 kt in 2030.
\begin{figure}[htp]
    \centering
    \includegraphics[width=15cm]{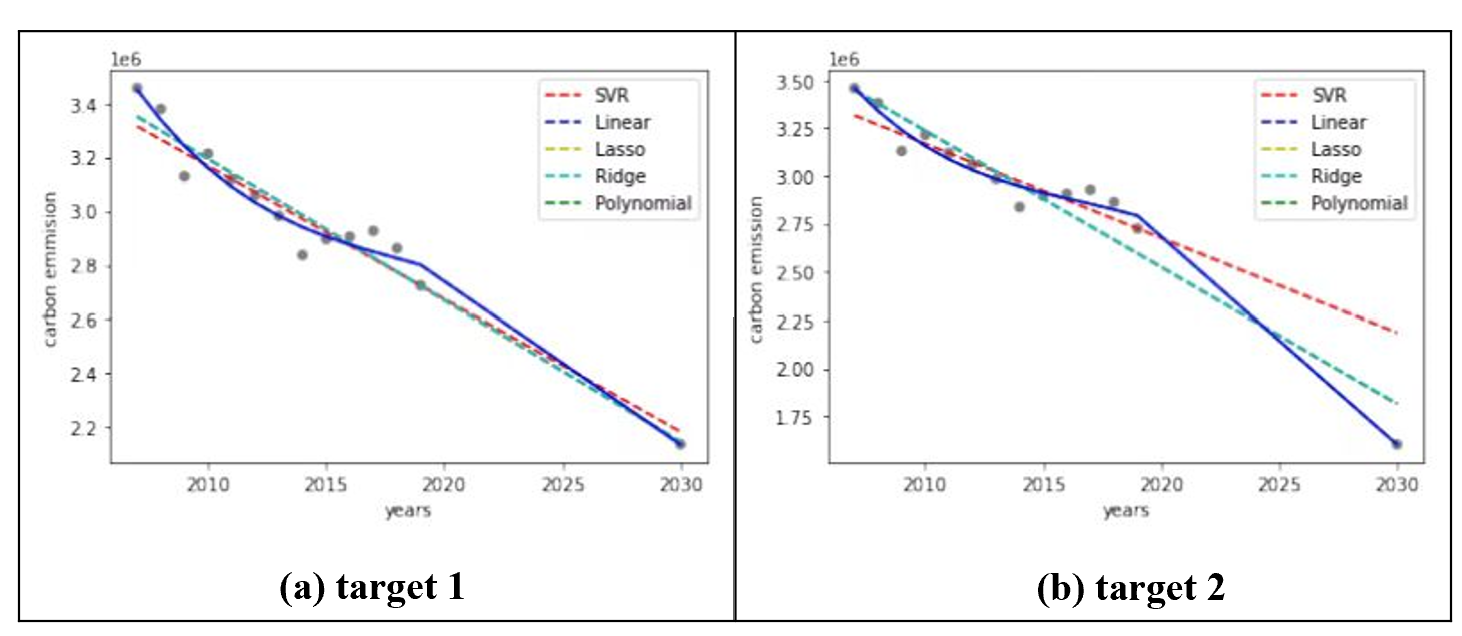}
    \caption{Forecast of carbon emission of different targets}
\end{figure}

\subsection{Forecast of carbon price}

   Using the prediction results of the above five inputs, we can roughly predict the carbon price through SVR. Since there are two different targets for the above carbon emission prediction, which lead to two different prediction results, so accordingly, there are two prediction results for carbon trading price. In Table 3, we can see the change data of carbon trading price under two targets in detail. Under target 1, the carbon price will reach about 114.72 euro in 2030, while under target 2, the carbon price can reach around 202.97 euro in 2030.
  \begin{center}[]
\begin{tabular}{c|c|c}
\hline
Year & Prediction (target 1) & Prediction (target 2) \\ \hline
2022 & 76.8426126            & 84.9632867            \\
2023 & 80.6052993            & 88.7259734            \\
2024 & 84.3703098            & 92.490984             \\
2025 & 88.1385497            & 96.2592238            \\
2026 & 91.9107871            & 100.031461            \\
2027 & 95.6874739            & 103.808148            \\
2028 & 99.4686523            & 107.589327            \\
2029 & 103.253998            & 111.374672            \\
2030 & 114.720389            & 202.968381            \\ \hline
\end{tabular}
\end{center}
\section{Comparison and Contrast with other models}
As shown in the graphs in the Result section, we have compared and contrast our SVR model with the other four models: Simple linear regression, Lasso regression, Ridge regression and polynomial regression. We use cross validation with leave-one-out method, which is equivalent to $K \  fold \  (n\ states = n)$. We found out that for all the predictions on the five factors, SVR has significantly better results in terms of cross validation score. See the table below.

\begin{center}
\begin{tabular}{||c c c c c c||} 
 \hline
 factors/models &  Oil & DAX & Coal & Gas \\ [0.5ex] 
 \hline\hline
 SVR & -201.93  & -2020572.79 & -692.31 &-424.67\\ 
 \hline
 Simple Linear & -479.63 & -2161172.47 &-970.38 &-514.67\\
 \hline
 Lasso & -479.62 & -2161173.04& -970.30 &-514.69\\
 \hline
 Ridge & -479.62 & -2161194.86 & -969.16 &-514.61\\
 \hline
 Polynomial & -479.63 & -2161172.47 &-970.37&-514.67\\ [1ex] 
 \hline
\end{tabular}
\end{center}
The scores represent the negative mean squared error, as we can see, SVR has the greatest negative mean squared error in each factor category. It means that SVR has the best generalization property.

\section{Discussion}

   By using the SVR tool, the future development trend of five monitored quantities and six sets of data can be regressed and visually presented in the figures.

From the results in the last section, it can be found that the two ways of prediction have different natures. We have already described how SVR works, and other traditional regression methods, they are analytical methods that use regression equations to model the relationship between one or more independent variables and dependent variables. However, the premise of using simple/regulized linear or polynomial regression is that there is indeed a linear/polynomial relationship between the independent variables and the dependent variable, otherwise, the results are very likely to be unable to explain the true situation. Moreover, regularized/simple linear or polynomial regression models are oversimplified and the actual situation is often much more complex, this approach has low accuracy and can lead to problems like overfitting/underfitting.

SVR provides a brand new method to solve the problem. It minimizes the distance of the most distant sample points in the hyperplane to obtain a function that minimizes the intra-class variance of all data. Meanwhile, the introduction of the kernel function enables SVR to deal with nonlinear and high-dimensional problems. In general, SVR offers a wider range of applications and improved accuracy than vanilla linear regression.
Certainly, it should not be ignored that SVR has many disadvantages in practical problems. For example, SVR performs poorly when large numbers of samples are encountered in high-dimensional situations, and it requires data to preprocess, which tests the user's ability to understand the data. To get more accurate predictions, we can also use other methods in the future, such as neural networks, a machine learning algorithm that has become very popular in recent years with excellent performance. It has the function of implementing any complex nonlinear mapping, but it is very complex, hard to guarantee the success of the process, requires rich input samples for training, and is difficult to understand.

\section{Conclusion}

This paper provides a way to predict European carbon prices using SVR and Paris climate accord until 2030. We utilize the adaptive power of SVR to predict the four factors that affect the carbon price most, then we forecast the carbon emissions from 2022 to 2030 using SVR and emission targets from Paris Accord. We also compare and contrast our model with other linear/polynomial regression models to illustrate the superiority of our model. 
  
  This research has the following implications: First incorporating the political variable into the 
  price predicting model can help improve the predicting accuracy, suggesting that investors should always consider the major political news/targets in decision-making and risk management. Moreover, when the European government sets a target for their carbon emission in the future, this paper can help them to forecast the carbon prices in the future based on their carbon emission targets.

  To our best knowledge, this paper is the first one to combine SVR and Paris accord to predict the carbon price, however, this study only has limited data to be used for training, and limited computational power to do the grid search for the best parameters of SVR, and only two policy targets used in the prediction. In the future, we hope that further research can develop and implement a more complex and robust model that improves the accuracy of forecasting.
  
\newpage
\bibliographystyle{IEEEtran}
\small\bibliography{reference}

\end{document}